\documentstyle[12pt,epsf]{article}

\advance\hoffset by -1.1truecm
\advance\voffset by -1.0truecm

\def\be{\begin{equation}}
\def\ee{\end{equation}}
\def\ba{\begin{eqnarray}}
\def\ea{\end{eqnarray}}
\def\v#1{\vert #1 \rangle}
\def\sp#1#2{\langle#1\vert#2\rangle}

\def\me#1#2#3{\langle #1 \vert #2 \vert #3 \rangle}

\newcommand{\N}{\mbox{I \hspace{-0.82em} N}}

\newcommand{\x}{{\bf x}}
\newcommand{\p}{{\bf p}}

\newcommand{\sn}{\smallskip\newline}
\newcommand{\mn}{\medskip\newline}
\newcommand{\bn}{\bigskip\newline}
\newcommand{\mbo}{{\mbox{ }}}

\def\Nc{{\cal{N}}}
\def\ml{\psi^{ml}}
\def\index{j}
\def\indone{\index}
\def\indtwo{ {j^{\prime}} }

\begin{document}

\title{Maximal localisation in the presence of minimal uncertainties
in positions and in momenta}
\author{Haye Hinrichsen\thanks{supported by Minerva foundation,
fehaye@wicc.weizmann.ac.il}\\ Department of Physics of Complex Systems\\
Weizmann Institute\\ Rehovot 76100, Israel\\ \\
Achim Kempf\thanks{supported by Studienstiftung des Deutschen
Volkes, BASF-Fellow, and Research Fellow of \ \ \ \ \ \ Corpus Christi
College in the Univ. of Cambridge, a.kempf@amtp.cam.ac.uk}\\
Department of Applied Mathematics \& Theoretical Physics\\
University of Cambridge\\
Cambridge CB3 9EW, U.K.}

\date{}
\maketitle

\begin{abstract}
Small corrections to the uncertainty relations,
with effects in the ultraviolet and/or infrared,
have been discussed in the context of string theory and quantum gravity.
Such corrections lead to small but finite minimal uncertainties in position
and/or momentum measurements.
It has been shown that these effects could indeed provide natural cutoffs in
quantum field theory. The corresponding underlying quantum
theoretical framework includes small `noncommutative geometric'
corrections to the canonical commutation relations.
In order to study the full implications on the concept of locality it
is crucial to find the physical states of then maximal
localisation. These states and their properties have been calculated
for the case with minimal uncertainties in positions only.
Here we extend this treatment, though still in one dimension, to the
general situation with minimal uncertainties both in positions and
in momenta.

\end{abstract}

\vskip-18truecm
\hskip11truecm
{\bf DAMTP/95-50}
\vskip18truecm

\newpage

\section{Introduction}
The short distance structure of conventional geometry
can be considered
experimentally confirmed up to the order of 1 TeV,
see e.g. \cite{khuri}.
In string theory and quantum gravity certain corrections to the short distance
structure and the uncertainty relations have been suggested to appear
at smaller scales (the latest at the Planck scale), see e.g.
\cite{townsend}-\cite{maggiore} and for a recent review \cite{garay}.

Here we continue a series of articles
\cite{ak-lmp-bf}-\cite{ak-gm-rm-prd} in which
are studied the quantum theoretical consequences of small corrections to the
canonical commutation relations
\be
[\x_i,\p_j] = i\hbar (\delta_{ij} + \alpha_{ijkl}\x_k\x_l +
\beta_{ijkl}\p_k\p_l +...)
\label{xp-ndim}
\ee
including the possibility that
also $[\x_i,\x_j] \ne 0$,\quad $[\p_i,\p_j]\ne 0$.
A crucial feature of
this `noncommutative geometric' ansatz,
which was first studied in \cite{ak-ixtapa}, is that
for appropriate matrices $\alpha$ and $\beta$, Eq.\ref{xp-ndim}
implies the existence of finite lower bounds to the determination of
positions and momenta. These bounds take the form of
finite minimal uncertainties
$\Delta x_0$ and $\Delta p_0$, obeyed by all physical states. In fact,
the approach
covers the case of those corrections to the uncertainty
relations which we mentioned above, see \cite{ak-jmp-ucr}.
\sn
A framework with a finite minimal uncertainty
$\Delta x_0$ can as well be understood to describe effectively nonpointlike
particles, than as describing a fuzzy space. As discussed in
\cite{ak-jmp-ucr}-\cite{ak-gm-rm-prd} the approach, with
appropriately adjusted scales, could have therefore more generally
a potential for an effective description of nonpointlike particles,
such as e.g. nucleons or quasi-particles in solids.
\sn
Analogously, on large scales
a minimal uncertainty $\Delta p_0$ may offer new possibilities to
describe situations where momentum cannot be precisely determined,
in particular on curved space \cite{ak-np3}.
\mn
Using the path integral formulation it has been shown in \cite{ak-np3}
that such noncommutative background geometries
can ultraviolet and  infrared regularise quantum field theories
in arbitrary dimensions through minimal uncertainties
$\Delta x_0, \Delta p_0$.
However, a complete analysis of the modified short distance structure,
and in particular the calculation of the states of maximal localisation,
has so far only been carried out for the special
case of the commutation relations
$[\x,\p] = i\hbar (1 + \beta \p^2)$, in \cite{ak-gm-rm-prd}. The reason is
that those cases are representation theoretically much easier to handle
in which either $\alpha$ or $\beta$ vanish, i.e. with minimal uncertainties
in either positions or in momenta only.
We now solve the more general, though still one-dimensional
problem, involving both minimal uncertainties
in positions and in momenta.
\mn
We define the associative Heisenberg
algebra $\cal{A}$ with corrections parametrised by small
constants $\alpha,\beta\ge0$
\be
[\x,\p] = i\hbar (1 + \alpha \x^2 + \beta \p^2)
\label{xp}
\ee
or, in a notation which will prove more convenient ($q\ge 1$)
\be
[\x,\p] = i\hbar \left(1 + (q^2-1)\left( \frac{\x^2}{4L^2} +
\frac{\p^2}{4K^2}\right)\right)
\label{xpq}
\ee
where the
constants $L,K$ carry units of length and momentum and are related by:
\be
4 KL=\hbar (1+q^2)
\ee
While the first correction term
contributes for large $\langle \x^2 \rangle = \langle \x\rangle^2
+ (\Delta x)^2$, which is the definition of the infrared,
the second correction term contributes for large $\langle \p^2 \rangle
= \langle \p\rangle^2 + (\Delta p)^2$, i.e. in the ultraviolet.
\sn
The corresponding uncertainty relation
\be
\Delta x \Delta p \ge \frac{\hbar}{2} \left(1+\alpha \left( (\Delta x)^2 +
\langle \x\rangle^2\right) +
\beta \left( (\Delta p)^2 + \langle \p\rangle^2\right)\right)
\ee
holds in all $*$-representations of the commutation relations and
reveals these
infrared and ultraviolet modifications as minimal uncertainties in positions
and momenta \cite{ak-jmp-ucr}:
\be
(\Delta x_{min})^2 \;=\; L^2 \, \frac{q^2-1}{q^2}
\left( 1 + (q^2-1) \left( \frac{\langle \x\rangle^2}{4L^2} +
\frac{\langle \p\rangle^2}{4K^2} \right) \right)
\ee
\be
(\Delta p_{min})^2 \;=\; K^2 \, \frac{q^2-1}{q^2}
\left( 1 + (q^2-1) \left( \frac{\langle \x\rangle^2}{4L^2} +
\frac{\langle \p\rangle^2}{4K^2} \right) \right)
\ee
In particular, for all physical states i.e. for all $\v{\psi} \in D$
with $D\subset H$
being any $*$-representation of the commutation relations of $\cal{A}$ in a
Hilbert space $H$, there are finite absolutly smallest uncertainties:
\be
(\Delta x_{\v{\psi}})
= \langle\psi\vert(\x-\langle \x\rangle)^2\vert\psi\rangle^{1/2} \ge
L\,\sqrt{1-q^{-2}}  \qquad \forall \quad \v{\psi}\in D
\label{nmu-x}
\ee
\be
(\Delta p_{\v{\psi}})
= \langle\psi\vert(\p-\langle \p\rangle)^2\vert\psi\rangle^{1/2} \ge
K\,\sqrt{1-q^{-2}}  \qquad \forall \quad \v{\psi}\in D
\label{nmu-p}
\ee
We will here only deal with the kinematical consequences of
possible corrections to the commutation relations.
Arbitrary systems can be considered and studies on
dynamical systems, including the calculation of the spectra of Hamiltonians
and integral kernels such as Green functions have been carried
out for example systems in \cite{ak-lmp-bf,ak-jmp-bf}.
Compare also with the features of the discretised quantum mechanics
studied e.g. in \cite{toy1,toy2,kw}.
A very interesting canonical field theoretical approach with a similar
motivation is focusing on generalising the uncertainty relations
among the coordinates \cite{doplicher 94/6}.

\section{Hilbert space representation}
A crucial consequence of Eqs.\ref{nmu-x},\ref{nmu-p} is that there are no
eigenvectors to $\x$ nor to $\p$ in any space of physical states
i.e. in any $*$-representation $D$ of the
generalised commutation relations. As is clear from the definition of
uncertainties, e.g.
$(\Delta x)^2_{\v{\psi}} = \langle \psi\vert(\x-\langle\psi\vert \x\vert
\psi \rangle)^2\vert \psi\rangle$, eigenvectors to $\x$ or
$\p$ could only have vanishing uncertainty in position or momentum.
In particular, the commutation relations of
$A$ no longer find spectral representations of $\x$ nor of $\p$.
\sn
In the situation of $\alpha=0$ (or $\beta=0$), i.e. with $\Delta p_0=0$
(or $\Delta x_0=0$)
there is still the momentum (or position) representation of
$\cal{A}$ available,
in which case the
maximal localisation states have been calculated in \cite{ak-gm-rm-prd}.
Let us now perform the analogous studies for the general case with
$\alpha, \beta >0$, where position and momentum space representations
are both ruled out.
\sn
To this end we use a Hilbert space
representation of $\cal{A}$ on a generalised Fock space.
The position and momentum operators can be represented as
\be
\x \;=\; L(a^\dagger + a)
\hspace{20mm}
\p \;=\; iK(a^\dagger - a)
\ee
where the $a$ and $a^\dagger$ obey generalised commutation relations
\be
a a^\dagger - q^2 a^\dagger a \;=\; 1
\ee
and act on the domain $D$ of physical states
$D:= \{\v{\psi}=\mbox{polynomial}(a^\dagger)\vert 0\rangle\}$ as:
\ba
a \,       \v{0} &=& 0 \\
a^\dagger  \v{n} &=& \sqrt{[n+1]} \, \v{n+1} \nonumber \\
a          \v{n} &=& \sqrt{[n]}   \, \v{n-1} \nonumber
\ea
where $[n]$ denotes the partial geometric sum or `$q$'- number
\be
[n] = \frac{q^{2n}-1}{q^2-1}
\ee
and where the $\v{n} := ([n]!)^{-1/2} (a^{\dagger})^n\vert 0\rangle,
\mbo n=1,...,\infty$ are orthonormalised
\be
\sp{n_1}{n_2} = \delta_{n_1,n_2}
\ee
and $D$ is analytic and dense in the Hilbert space $H=l^2$.
\sn
While $\x$ and $\p$ ordinarily are
essentially self-adjoint, they are now merely symmetric, which is
sufficient to insure that all expectation values are real.
The deficiency indices of $\x$ and $\p$ are $(1,1)$, implying the existence
of 1-parameter families of self-adjoint extensions. While ordinarily
self-adjoint extensions, e.g. for a particle in a box, need to
and can be fixed,
there is now the subtle effect of the self-adjoint extensions not being
on common domains, which prevents the diagonalisation of $\x$ or $\p$
on physical states, as can also be understood through the uncertainty
relations.
For the full functional analytical details see \cite{ak-jmp-ucr}, where
these structures have first been found. We will come back to these
functional analytical studies in Sec.\ref{inv} where we will
explicitly calculate the diagonalisations in $H$.
They are of use for
the calculation of inverses of $\x$ and $\p$, which are
not only needed to describe certain quantum mechanical potentials,
but ultimately also to invert kinetic terms
e.g. of the form $\p^2-m^2$ to obtain propagators from the field theoretical
path integral, see \cite{ak-np3}.

\section{Maximal localisation states}
The absence of eigenvectors of $\x$ or $\p$
in all $*$-representations $D$ of the
commutation relations physically implies the absence of absolute
localisability in position or momentum i.e. there are no physical states
that would have $\Delta x =0$ or $\Delta p=0$. More precisely,
the uncertainty relation, holding in all $D$, implies
a `minimal uncertainty gap':
\be
\exists\!\!\!\!\!\!\not \mbo \mbo
\v{\psi} \in D: \mbo \Delta x_{\v{\psi}} < \Delta x_0 \qquad \mbox{and}
\qquad \exists\!\!\!\!\!\!\not \mbo \mbo
\v{\psi} \in D: \mbo \Delta p_{\v{\psi}} < \Delta p_0\;.
\label{gap}
\ee
The state of maximal localisation in position $\v{\ml_x}$ with
given position expectation $x$ and vanishing momentum expectation,
is defined through
\be
\label{MaxLocDef}
\me{\ml_x}{\x}{\ml_x} = x\;,
\hspace{15mm}
\me{\ml_x}{\p}{\ml_x} = 0\;,
\hspace{15mm}
(\Delta x)_{\v{\ml_x}}  = \Delta x_{min}\;.
\ee
Explicitly the minimal uncertainty in position then reads
\be
\label{DeltaX}
(\Delta x)_{\v{\ml_x}}^2 = L^2 \, \frac{q^2-1}{q^2}
\left( 1 + (q^2-1) \frac{\langle\x\rangle^2}{4L^2} \right)
\ee
with the corresponding (now not infinite) uncertainty in momentum:
\be
\label{DeltaP}
(\Delta p)_{\v{\ml_x}}^2 = K^2 \, \frac{(q^2+1)^2}{q^2(q^2-1)}
\left( 1 + (q^2-1) \frac{\langle\x\rangle^2}{4L^2} \right) \;.
\ee
We focus on maximal localisation in $x$, the case of maximal localisation
in $p$ is fully analogous.
\sn
As shown in \cite{ak-gm-rm-prd} a state of maximal localisation
is determined by the equation
\be
\biggl((\x-\langle\x\rangle) + i \alpha (\p-\langle\p\rangle)
\biggr)\,\v{\ml_x} \;=\; 0
\ee
where $\alpha = \Delta x/\Delta p$.
Inserting Eqs.\ref{DeltaX},\ref{DeltaP} we obtain
\be
\alpha \;=\; \frac{L(q^2-1)}{K(q^2+1)}
\ee
so that the condition reads:
\be
\label{MaxLocEq}
\left( \frac{q^2+1}{L} \, (\x-\langle \x\rangle) \,+\,
		 i \frac{q^2-1}{K} \, \p \right)
\v{\ml_x} \;=\; 0\;.
\ee
\subsection{Maximal localisation states in the Fock basis}
In order to explicitly calculate
those states that realise the now maximally possible localisation
we expand the $\v{\ml_x}$ in the Fock basis
\be
\label{BFRep}
\v{\ml_x} \; := \; \frac{1}{\Nc(x)} \,
\sum_{n=0}^\infty q^{-3n/2} \, c_n(x) \, \v{n}
\ee
where the $c_n(x)$ are real coefficients and
\be
\label{NormDefinition}
\Nc(x) \; := \; \sum_{n=0}^\infty \, q^{-3n} \, c_n^2(x)
\ee
is a normalisation factor
(the inserted factors $q^{-3n/2}$ will be convenient later).\newline
The condition for maximal localisation Eq.\ref{MaxLocEq}
reads in the Fock representation:
\be
\label{MaxLocEqBF}
\left( (q^2+1)(a^\dagger+a-\frac{x}{L}) \,-\,
	  (q^2-1)(a^\dagger-a) \right)
\v{\ml_x} \;=\; 0\;.
\ee
Inserting the ansatz Eq.\ref{BFRep} we are led to
the recursion relation
\be
\frac{q+q^{-1}}{2L} \, x \, c_n(x) \;=\;
\sqrt{q^{-1}[n+1]}\,c_{n+1}(x) \,+\,
\sqrt{q[n]}       \,c_{n-1}(x)\;.
\label{reca}
\ee
Together with
\be
c_{-1}(x) = 0
\hspace{7mm} \mbox{and} \hspace{7mm}
c_0(x) = 1
\ee
the coefficients $c_n(x)$ are therefore determined as
polynomials of degree $n$ in $x$.
\subsection{Relation to continuous $q$-Hermite polynomials}
The coefficients $c_n(x)$ are related to
the so-called continuous $q$-Hermite polynomials. An excellent review
on these and other $q$-orthogonal polynomials is \cite{ReportDelft}.
\sn
We use the notation of shifted $q$-factorials \cite{ReportDelft}
\be
\label{QFactorials}
(a;q^2)_n \; := \; \prod_{k=0}^{n-1} \, (1-a q^{2k})
\ee
which obey the identity
\be
\label{Qinverse}
(a;q^2)_n \;=\; (-a)^n\, q^{n(n-1)} \, (a^{-1};q^{-2})_n\,.
\ee
Furthermore we define for later convenience
\be
\label{XJFunction}
\index(x) := \frac{\mbox{arcsinh}(\omega x)}{\ln q}\;,
\hspace{20mm}
x(\index) = \frac{q^\index-q^{-\index}}{2\omega}
\ee
where
\be
\label{DefOmega}
\omega \; :=\; \frac{1}{4L}\,(q+q^{-1})\sqrt{q^2-1}\,.
\ee
The continuous $q$-Hermite polynomials $H_n(z|q^2)$
are defined through
\be
H_{-1}(z|q^2)=0,
\qquad
H_{0}(z|q^2)=1
\ee
and the recurrence relation, see Ref.\cite{ReportDelft}:
\be
2 z H_n(z|q^2) \;=\; H_{n+1}(z|q^2) \,+\, (1-q^{2n}) H_{n-1}(z|q^2)\;.
\ee
It is not difficult to check that this recursion relation can be brought
into the form of the recursion relation Eq.\ref{reca} for the
coefficients $c_n(x)$, by expressing them in terms of the $H_n(z\vert q^2)$
as:
\be
\label{Correspondence}
c_n(x) \;=\; \sqrt{\frac{q^n}{[n]!\,(q^2-1)^n}} \, i^{-n}
H_n\Bigl(i \omega x \,|\, q^2\Bigr)\;.
\ee
As shown in \cite{ReportDelft}, the continuous
$q$-Hermite polynomials $H_n(z|q^2)$ can be written as
\be
\label{BinominalExpression}
H_n(z|q^2) \;=\; \sum_{k=0}^n \,
\biggl(\hspace{-1mm} \begin{array}{c} n \\ k
\end{array}\hspace{-1mm}  \biggr)_{q^2}
e^{i(n-2k)\theta} \,,
\hspace{15mm}
z=\cos\theta
\ee
with the $q$-binomial coefficients:
\be
\biggl(\hspace{-1mm}  \begin{array}{c} n \\
k \end{array}\hspace{-1mm}  \biggr)_{q^2}=\;
\frac{(q^2;q^2)_n}{(q^2;q^2)_k\,(q^2;q^2)_{n-k}}\;.
\ee
Inserting Eq.\ref{BinominalExpression} into Eq.\ref{Correspondence} and
replacing $[n]!$ by
\be
[n]! \;=\,
\frac{(-)^n (q^2;q^2)_n}{(q^2-1)^n} \;=\;
\frac{q^{n^2} (q^{-2};q^{-2})_n}{(q-q^{-1})^n}
\ee
yields
\be
c_n(x) \;=\;
\frac{1}{\sqrt{q^{n^2} (q^{-2}; q^{-2})_n}} \, i^{-n} \,
\sum_{k=0}^n \,
\biggl(\hspace{-1mm} \begin{array}{c} n \\ k
\end{array}\hspace{-1mm}  \biggr)_{q^2}
e^{i(n-2k)\theta} \,,
\hspace{15mm}
i \omega x=\cos\theta\;.
\ee
Because of $i \omega x = \frac12 (q^{\index(x)}- q^{-\index(x)})$
we may also write $e^{i \theta} = i\,q^{\index(x)}$ and therefore obtain
the following exact expression for the coefficients $c_n(x)$:
\be
\label{Cexplicit}
c_n(x) \;=\;
\frac{1}{\sqrt{q^{n^2}\,(q^{-2};q^{-2})_n}}
\sum_{k=0}^n\,
\biggl(\hspace{-1mm}  \begin{array}{c} n \\
 k \end{array}\hspace{-1mm}  \biggr)_{q^2} \,
(-)^{k} \, q^{(n-2k)\,\index(x)}\;.
\label{a3}
\ee
We derive further useful properties of the $c_n(x)$. \bf
\newpage
Classical limit \rm \\[2mm]
%
For $q \rightarrow 1$ the recursion relation Eq.\ref{reca} reduces to
\be
\frac{x}{L} c_n(x) \;=\; \sqrt{n+1}\,c_{n+1}(x) \,+\,
\sqrt{n} \, c_{n-1}(x)\;.
\ee
By substituting
$x=L\sqrt{2}\,z$ and $H_n(z) = \sqrt{n!\,2^n}\,c_n(x)$ we obtain
$H_0(z)=1$ and
\be
2z\,H_n(z) \;=\; H_{n+1}(z) + 2n\,H_{n-1}(z)
\ee
which is the defining recursion relation for classical Hermite polynomials
$H_n(z)$. Thus the classical limit of the polynomials $c_n(x)$ is given by
\be
\lim_{q \rightarrow 1} c_n(x) \;=\;
\frac{1}{\sqrt{n!\,2^n}}\,
H_n\Bigl(\frac{x}{L \sqrt{2}}\Bigr)\;.
\ee
\mn
{\bf Representation by the formula of Rodriguez}\\[2mm]
%
As a short notation we write $x(\index)$ as $x_\index$. Then,
introducing the $q$-difference operator
\be
D\,f(x_\index) \;=\; \frac{f(x_{\index+1})-f(x_{\index-1})}
			  {x_{\index+1}-x_{\index-1}}
\label{R1}
\ee
the polynomials $c_n(x_\index)$ can be expressed as
\be
\label{Rodriguez}
c_n(x_\index) \;=\; \frac{(-)^n}{\kappa_n}\,
q^{\index^2} \, D^n \, q^{-\index^2}
\ee
where
\be
\kappa_n \;=\; \sqrt{q^{-n^2}\, [n]!}\,
\biggl( \frac{q+q^{-1}}{2L} \biggr)^n\;.
\ee
Eq.\ref{Rodriguez} generalises the formula of Rodriguez
$H_n(x) = (-)^n e^{x^2} \frac{d^n}{dx^n} e^{-x^2}$
for classical Hermite polynomials. A proof for Eq.\ref{Rodriguez}
is outlined in appendix A. \\[5mm]
{\bf $q$-difference equation}\\[2mm]
%
The generalised formula of Rodriguez Eq.\ref{Rodriguez} implies that
\be
\label{DifferenceEquation}
c_n(x_{\index+1})-c_n(x_{\index-1}) \;=\;
\sqrt{q(1-q^{-2n})}\, (q^\index+q^{-\index})\,
c_{n-1}(x_\index)
\ee
which generalises the differentiation rule
$\frac{d}{dx} H_n(x) = n H_{n-1}(x)$ for classical Hermite polynomials.
In order to prove this equation, we rewrite its l.h.s. using
Eqs.\ref{R1},\ref{Rodriguez}
\be
c_n(x_{\index+1})-c_n(x_{\index-1}) \;=\;
(x_{\index+1}-x_{\index-1}) \, D \, c_n(x_\index) \;=\;
(x_{\index+1}-x_{\index-1}) \, D\, \frac{(-)^n}{\kappa_n}\,
q^{\index^2} \, D^n \, q^{-\index^2}\,.
\ee
Carrying out the first differentiation on the r.h.s. of this formula
(c.f. Eq.\ref{ProductRule}), one
obtains a linear combination of the polynomials
$c_{n}(x_\index)$ and $c_{n+1}(x_\index)$
which in turn can be expressed through
the recurrence relation Eq.\ref{reca} in terms of $c_{n-1}(x_\index)$.
\bn
It can also be shown by induction that Eq.\ref{DifferenceEquation}
implies the following $q$-difference equation for the
polynomials $c_n(x)$
\be
\label{QDiff}
q^\index c_n(x_{\index-1}) + q^{-\index} c_n(x_{\index+1}) \;=\;
q^{-n}\,(q^\index+q^{-\index})\,c_n(x_\index)
\ee
which corresponds to the differential equation
$2xH'_n(x) - H''_n(x) = 2nH_n(x)$
for classical Hermite polynomials.\\[5mm]
{\bf Orthogonality}\\[2mm]
%
The polynomials $c_n(x)$ obey the orthogonality relation
\be
\label{OrthoRel}
\sum_{\index=-\infty}^\infty \,
(x_{2\index+\kappa+1}-x_{2\index+\kappa-1})\,
q^{-(2\index+\kappa)^2} \,
c_m(x_{2\index+\kappa}) c_n(x_{2\index+\kappa}) \;=\;
N_\kappa\, q^n\, \delta_{m,n}
\ee
where
\be
N_\kappa \;=\;
\sum_{\index=-\infty}^\infty \,
(x_{2\index+\kappa+1}-x_{2\index+\kappa-1})\, q^{-(2\index+\kappa)^2}\;.
\ee
The parameter $0 \leq \kappa \leq 2$ can be choosen arbitrarily and
fixes a family of positions occurring in the sum.
\\
Eq.\ref{OrthoRel} can be proved as follows.
The case $m=n=0$ is trivial. For $n=0$ and
$m>0$ one can show that the l.h.s. of Eq.\ref{OrthoRel} is equal to
\be
\sum_{\index=-\infty}^\infty \,
(x_{2\index+\kappa+1}-x_{2\index+\kappa-1})\,D^m\,
q^{-(2\index+\kappa)^2} \;=\;
\left. D^{m-1} \, q^{-j^2} \right|_{j=-\infty}^{j=+\infty} \;=\; 0\;.
\ee
Keeping $m$ fixed, a further induction for $n>0$ completes the proof.
\\[5mm]
{\bf Generating function}\\[2mm]
%
A generating function of the polynomials $c_n(x)$ is
\be
(t\,q^{-\index(x)}; q^{-2})_\infty (-t\,q^{\index(x)};q^{-2})_\infty \;=\;
\sum_{n=0}^\infty \,
\frac{c_n(x)}{\sqrt{q^{n(n-2)}\,(q^{-2};q^{-2})_n}} \, t^n\;.
\ee
In order to verify this expression, we use the
$q$-difference equation Eq.\ref{QDiff} and get
\ba
&&q^j\,(t\,q^{-\index+1}; q^{-2})_\infty \,
(-t\,q^{\index-1};q^{-2})_\infty\;+\;
q^{-j}\,(t\,q^{-\index-1}; q^{-2})_\infty \,
(-t\,q^{\index+1};q^{-2})_\infty
\nonumber \\&& \;=\;(q^j+q^{-j})\,
(t\,q^{-\index-1}; q^{-2})_\infty \,
(-t\,q^{\index-1};q^{-2})_\infty
\ea
which in turn can be proved by inserting
the definition of the $q$-factorials (c.f. Eq.\ref{QFactorials}).
\section{Quasi- position and momentum wave functions}
Generally, all information on position and momentum is contained
in the matrix elements of the position and momentum operators, and
matrix elements can of course be calculated in arbitrary bases, such
as also the Fock basis. Ordinarily, the position and the momentum
information content of a state $\v{\phi}$ of the particle is easily
obtained by writing the state as a position or momentum space
wave function $\phi(x) = \sp{x}{\phi}$ or $\phi(p) = \sp{p}{\phi}$,
which is to project onto
position or momentum eigenstates, i.e. to project onto
states of maximal localisation in $x$ or $p$.
\sn
In the new setting we can now project arbitrary states $\v{\phi}$
onto the states which realise the maximally possible
localisation in position (or in momentum), which are given by
Eqs.\ref{MaxLocDef},\ref{BFRep},\ref{a3}. We call the collection of
these projections the quasi-position wavefunction $\phi(x)$ of $\v{\phi}$:
\be
\phi(x) := \langle \psi^{ml}_x\vert\phi\rangle
\ee
Here $\phi(x)$ yields
the probability amplitude for finding the particle in a state of maximal
localisation around the position $x$ with vanishing momentum expectation.
As is easily seen from Eqs.\ref{MaxLocDef},
\ref{MaxLocEq},\ref{MaxLocEqBF}
the generalisation to
arbitrary momentum expectations is straightforward. The framework for
quasi-momentum wave functions
\be
\phi(p) := \langle \psi^{ml}_p\vert\phi\rangle
\ee
is analogous with $\phi(p)$ being the probability amplitude for
finding the particle in a state of maximal localisation in its momentum,
with the momentum expectation $p$ and vanishing position expectation
(again the definition may easily be generalised to include arbitrary position
expectations).
\sn
Aiming at the calculation of examples of quasi-wave functions, we need
to complete our studies on the maximal localisation states
by calculating their norm and scalar product.
To this end an important technical tool will be the Christoffel Darboux
theorem, for the application of which we will need the limiting cases
of the coefficients $c_n(x)$ of the maximal localisation states.

\subsection{Limits of $(-1)^n c_{2n}(x)$ and $(-1)^n
c_{2n+1}(x)$ for $n \rightarrow \infty$}
As we prove in appendix B, the polynomials $c_n(x)$, for all odd
and for all even $n$ have
the nontrivial property that their limit for $n \rightarrow \infty$
exists. Denoting again $x_\index := x(\index) $
these limits are:
\ba
\label{Cplus}
c^+(x_\index) &=& \lim_{m \rightarrow \infty} \, (-)^m \, c_{2m}(x_\index)
\hspace{4mm}
\;=\; A\,q^{\index^2/2}\,\theta_2 (\frac{\pi \index}{2},\lambda) \\
\label{Cminus}
c^-(x_\index) &=& \lim_{m \rightarrow \infty} \, (-)^m \, c_{2m+1}(x_\index)
\;=\; A\,q^{\index^2/2}\,\theta_1 (\frac{\pi \index}{2},\lambda)
\ea
where $\theta_i(z,\lambda)$ are the Jacobi-, or elliptic $\theta$-functions
defined as
\ba
\label{Theta1}
\theta_1(z,\lambda) &:=& 2 \lambda^{1/4}
\sum_{n=0}^\infty (-)^n \lambda^{n(n+1)} \, \sin((2n+1)z) \\
\theta_2(z,\lambda) &:=& 2 \lambda^{1/4}
\label{Theta2}
\sum_{n=0}^\infty \lambda^{n(n+1)} \, \cos((2n+1)z)
\ea
and where in Eqs.\ref{Cplus},\ref{Cminus}
the constants $\lambda$ and $A$ are defined as
\be
\label{ConstLambda}
\lambda := e^{\frac{-\pi^2}{2 \ln q}}
\ee
and
\be
\label{ConstA}
A^2 \; :=\; \frac{\pi}{2 \,(q^{-2};q^{-2})_\infty^3\,\ln q }
\;=\; \frac{2}{q^\frac14 \, \theta_2(0,q^{-1})\,
		  \theta_2^2(0,\lambda)}\;.
\ee
\subsection{Normalisation and scalar product of
maximal localisation states}
In order to evaluate the
scalar product of two maximally localised states
\be
\sp{\ml_{x}}{\ml_{x^{\prime}} } \;=\;
\frac{1}{\sqrt{\Nc(x)\Nc(x^\prime)}}
\sum_{n=0}^\infty\,
q^{-3n}\,c_n(x)\,c_n(x^\prime)
\ee
the $q$-difference equation Eq.\ref{QDiff} can be used to rewrite
this expression as
\be
\label{QuasiScalarProductDef}
\sp{\ml_{x}}{\ml_{x^\prime}} \;=\;
\frac{q^{\indone+\indtwo}f_{\indone-1,\indtwo-1} \,+\,
	 q^{\indone-\indtwo}f_{\indone-1,\indtwo+1} \,+\,
	 q^{\indtwo-\indone}f_{\indone+1,\indtwo-1} \,+\,
	 q^{-\indone-\indtwo}f_{\indone+1,\indtwo+1}}
    {(q^\indone+q^{-\indone})\,
	(q^\indtwo+q^{-\indtwo})\,
	\sqrt{\Nc(x)\Nc(x^\prime)}}
\ee
where we defined
\be
f_{\indone,\indtwo} \; := \; \sum_{n=0}^\infty\,
q^{-n}\,c_n(x)\,c_n(x^\prime)
\ee
and where we abbreviated $\index := \index(x)$
and $\index^\prime := \index(x^\prime)$.
We can compute $f_{\indone,\indtwo}$ by applying the
Christoffel-Darboux \cite{christoffel-darboux} theorem
\be
\label{Christoffel}
\sum_{n=0}^m\,
q^{-n}\,c_n(x)\,c_n(x^\prime) \;=\;
\frac{2L\sqrt{[m+1]} }{q^{m+\frac12}(q+q^{-1})}\,\,\,
\frac{c_{m+1}(x)c_m(x^\prime)\,-\,c_m(x)c_{m+1}(x^\prime)}{x-x^\prime}
\ee
which can be proved as follows. For $m>0$
(the case $m=0$ is trivial) we use the recursion relation (c.f. Eq. \ref{reca})
\be
c_{m+1}(x) \;=\; \sqrt{\frac{q}{[m+1]}} \,
\Bigl( \frac{q+q^{-1}}{2L}\,x\,c_m(x) \,-\, \sqrt{q[m]} \, c_{m-1}(x) \Bigr)
\ee
in order to replace $c_{m+1}(x)$ and $c_{m+1}(x^\prime)$
on the r.h.s. of Eq.\ref{Christoffel} which then takes the form
\be
\mbox{r.h.s.} \;=\; q^{-m}\,\,c_m(x)\,c_m(x^\prime) \,-\,
\frac{2L\sqrt{[m]} }{q^{m-\frac12}(q+q^{-1})}\,\,
\frac{c_{m-1}(x)c_m(x^\prime)\,-\,c_m(x)c_{m-1}(x^\prime)}{x-x^\prime}
\ee
so that Eq.\ref{Christoffel} follows by induction.
\sn
Since the polynomials $c_m(x)$ have well defined limits
as $m$ goes to infinity, the Christoffel-Darboux theorem
implies that the expression $f_{j,j^\prime}$ is given by
\be
\label{NormalScalarProduct}
f_{\indone,\indtwo}  \;=\;
\sum_{n=0}^\infty\,
q^{-n}\,c_n(x)\,c_n(x^\prime) \;=\;
\frac{\sqrt{q}}{2\omega}\,\,
\frac{c^-(x)c^+(x^\prime)-c^+(x)c^-(x^\prime)}
	{x-x^\prime}\;.
\ee
Inserting Eqs.\ref{Cplus},\ref{Cminus} yields
\be
f_{\indone,\indtwo}  \;=\;
\frac{2\,A^2\,q^{\frac12(\indone^2+\indtwo^2+1)}}
	 {q^\indone-q^{-\indone} - q^\indtwo + q^{-\indtwo}}\,\,
\theta_1\Bigl(\frac{\pi}{2}(\indone-\indtwo),\lambda^2\Bigr)\,\,
\theta_4\Bigl(\frac{\pi}{2}(\indone+\indtwo),\lambda^2\Bigr)
\label{a60}
\ee
with the definition of $\theta_4$ being
\be
\theta_4(z,\lambda^2) \; :=\; 1\,+\,2\sum_{n=1}^\infty\,
(-)^n\,\lambda^{2n^2}\,\cos(2nz)\;.
\ee
In the limit $x \rightarrow x^\prime$ Eq.\ref{a60} reduces to
\be
\label{NormalNorm}
f_{\index,\index}  \;=\;
\sum_{n=0}^\infty\,
q^{-n}\,c_n^2(x) \;=\;
\frac{q^{\index^2+\frac14} \,\theta_2(0,q^{-1})}
	{(q^{\index}+q^{-\index})\,\theta_4(0,\lambda^2)}
\, \theta_4(\pi \index,\lambda^2)\;.
\ee
Inserting Eq.\ref{NormalScalarProduct} into
Eq.\ref{QuasiScalarProductDef} we eventually obtain an exact
expression for the scalar product of two quasi-position states:
\be
\label{QuasiScalarProduct}
\sp{\ml_{x}}{\ml_{x^\prime}} \;=\;
\frac{2A^2\, q^{\frac12(\indone^2+\indtwo^2+1)}
	 (q^2-1)^2 (1+q^{-2})}
	{\sqrt{\Nc(x)\Nc(x^\prime)}\,G^0_{\indone,\indtwo}\,
	 G^1_{\indone,\indtwo}\,G^{-1}_{\indone,\indtwo}}
	\,\theta_1(\frac{\pi}{2}(\indtwo-\indone),\lambda^2)
	\,\theta_4(\frac{\pi}{2}(\indone+\indtwo),\lambda^2)
\ee
where
\be
G^s_{\indone,\indtwo} \;=\;
(q^{\frac12(\indone-\indtwo)+s}-q^{-\frac12(\indone-\indtwo)-s})
(q^{\frac12(\indone+\indtwo)+s}+q^{-\frac12(\indone+\indtwo)-s})\,.
\ee
Note that the poles in the denominator of Eq.\ref{QuasiScalarProduct}
are cancelled by the zeros of the $\theta_1$-function.
The limit $x \rightarrow x^\prime$ yields the
norm (Eq.\ref{NormDefinition})
\be
\label{QuasiNorm}
\Nc(x) \;=\;
\frac{2\,q^{\index^2} \,
	 (q^2+1) \, \theta_4(\pi \index, \lambda^2)}
	{A^2(q^{\index}+q^{-\index})\,(q^{\index+1}+q^{-\index-1})\,
	 (q^{\index-1}+q^{-\index+1})\,
	 \theta_2^2(0,\lambda)\,\theta_4(0,\lambda^2)}\;.
\ee
\subsection{Example: The quasi-position wave function of $\v{\ml_0}$.}
As an example we draw the graph of the quasi-position wave function
$\phi(x)$ for the state $\v{\phi}$ that describes maximal localisation
around $x=0$ i.e. for $\v{\phi}:= \v{\psi^{ml}_x}$, i.e. with
\be
\phi(x) = \sp{\psi^{ml}_x}{\psi^{ml}_0}
\ee
\vskip3truecm
\epsfxsize=5in \centerline{\epsfbox{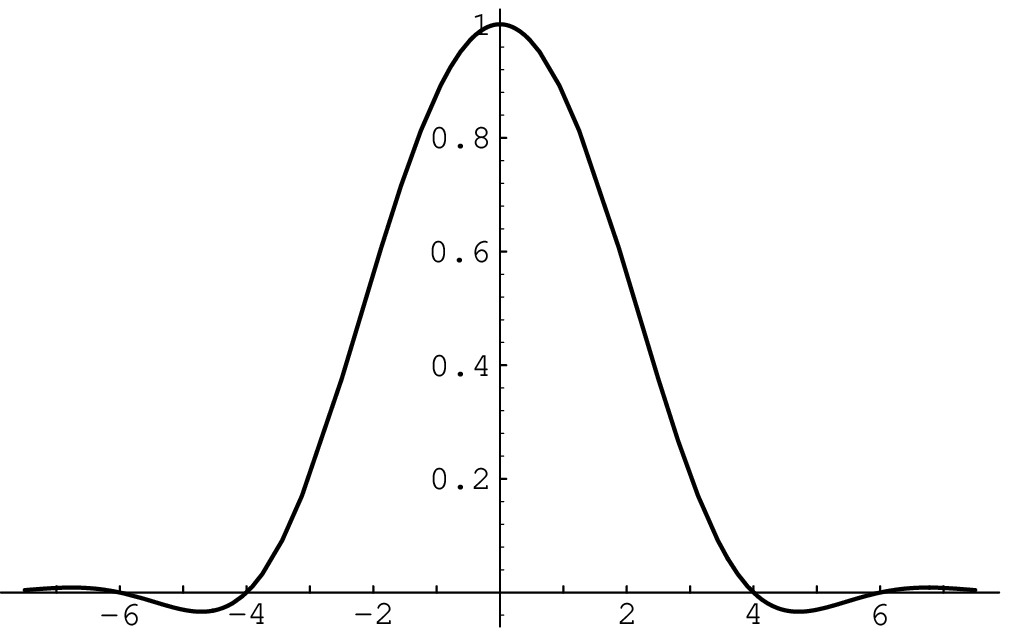}}
\centerline{{ \small \it Fig. 1: Quasi-position wave function $\phi(x)$ for
$\v{\phi} := \v{\psi^{ml}_0}$
for $q=1.5$ drawn over $j(x)$ }}
\vskip0.8truecm

\noindent
The analytic form of the wave function is given in
Eq.\ref{QuasiScalarProduct}. For the width of the main peak note that
the graph shows the overlap of pairs of localisation states,
each with its finite position uncertainty.

We have thus generalised the treatment of \cite{ak-gm-rm-prd} where
the corresponding graph was calculated and drawn
for the special case without a
minimal momentum uncertainty ($\alpha = 0, \beta > 0$).

\section{Approximations}
For potential applications of the formalism the parameters $\alpha$ and
$\beta$ in Eq.\ref{xp} can be assumed small in which case useful
simplifications hold.
\sn
In the notation of Eq.\ref{xpq} this is the case when $q \rightarrow 1$.
Then $\lambda \rightarrow 0$ and the $\theta$-functions
in Eqs.\ref{Cplus},\ref{Cminus} can be approximated by
$\theta_1(\frac{\pi \index}{2},\lambda) \approx 2 \lambda^{1/4} \sin
\frac{\pi \index}{2}$ and
$\theta_2(\frac{\pi \index}{2},\lambda) \approx 2 \lambda^{1/4} \cos
\frac{\pi \index}{2}$. This implies that
\ba
\label{Capprox}
c^+(x_\index) &\approx& \tilde{c}^+(x_\index) \; :=\;
B \, q^{\index^2/2} \,
\cos \frac{\pi \index}{2} \\
c^-(x_\index) &\approx& \tilde{c}^-(x_\index) \; :=\;
B \, q^{\index^2/2} \,
\sin \frac{\pi \index}{2} \nonumber
\ea
where
\be
\label{Bdef}
B^4 \;=\; \frac{4 \,\ln q}{\pi \sqrt{q}}\;.
\ee
The relative error of this approximation is shown in the
following table:

\def\Nexta   {   \hline        && \\[-5mm]}
\def\Next    {\\ \hline        && \\[-4mm]}
\def\Nextb   {\\ \hline \hline && \\[-4mm]}
\def\NextEnd {\\ \hline}
\begin{table}[ht]
\begin{center}
\begin{tabular}{|c||c|c|c|c|}
\Nexta
$q$ & 1.2 & 1.5 & 2 & 5
\Nextb
$|1-\frac{\tilde{c}^\pm(x)}{c^\pm(x)}|$
& $< 5 \times 10^{-15}$
& $< 1 \times 10^{-10}$
& $< 2 \times 10^{-6}$
& $< 7 \times 10^{-3}$
\NextEnd
\end{tabular}
\end{center}
\end{table}
\noindent
Using Eq.\ref{Capprox} we can give approximations of the scalar product
for $q$ close to $1$:
\be
\label{ApproxScalarProduct}
\sp{\ml_{x}}{\ml_{x^\prime}} \;\approx\;
\frac{B^2\, q^{\frac12(\indone^2+\indtwo^2+1)}
	 (q^2-1)^2 (1+q^{-2})}
	{\sqrt{\tilde{\Nc}(x)
	    \tilde{\Nc}(x^\prime)}\,G^0_{\indone,\indtwo}\,
	 G^1_{\indone,\indtwo}\,G^{-1}_{\indone,\indtwo}}
	\,\sin\frac{\pi}{2}(\indtwo-\indone)
\ee
where
\be
\label{NcNorm}
\Nc(x) \;\approx\; \tilde{\Nc}(x) \; :=\;
\frac{2\,q^{\index^2} \,(q^2+1)}
	{B^2(q^{\index}+q^{-\index})\,(q^{\index+1}+q^{-\index-1})\,
	 (q^{\index-1}+q^{-\index+1})}\;.
\ee
E.g. for $1<q<1.2$ the relative error of this approximation is less
than $10^{-14}$.
\bn
It is interesting to consider also the limiting case where
\be
q \rightarrow 1, \qquad K(q) := \sqrt{\frac{q^2-1}{4 \beta}}, \qquad
L(q) = \hbar \frac{q^2+1}{2} \sqrt{\frac{\beta}{q^2-1}}\,.
\label{limits}
\ee
In this limit the commutation relations Eq.\ref{xpq} turn into
the relations Eq.\ref{xp} with $\beta$ finite but $\alpha=0$, which is
the special case considered in \cite{ak-gm-rm-prd}.
There is then only a minimal uncertainty in position
and no minimal uncertainty in momentum. As can be shown easily, the
limit $q \rightarrow 1$ of the scalar product
Eqs.\ref{ApproxScalarProduct}-\ref{NcNorm} is given by
\be
\lim_{q \rightarrow 1}\,\sp{\ml_{x^\prime}}{\ml_{x}} \;=\;
\frac{\sin \frac{\pi}{2}(j^\prime-j)}
     {\pi\,(\frac{j-j^\prime}{2})\,(\frac{j-j^\prime}{2}+1)\,
           (\frac{j-j^\prime}{2}-1)}\,.
\ee
In the limit given by Eqs.\ref{limits}, $x$ and $j$ are related linearly
through $x_j = x(j) = \hbar \sqrt{\beta} \, j$. We thus obtain the limiting
expression for the scalar product:
\be
\langle \psi^{ml}_{x^{\prime}} \vert \psi^{ml}_{x} \rangle =
 \frac{1}{\pi} \left( \frac{{x}-{x}^{\prime}}{2\hbar\sqrt{\beta}}
 - \left(\frac{{x}-{x}^{\prime}}{2\hbar \sqrt{\beta}}\right)^3\right)^{-1}
 \sin\left(\frac{{x}-{x}^{\prime}}{2\hbar\sqrt{\beta}}\pi\right)
\ee
This result coincides with the expression found in \cite{ak-gm-rm-prd},
thus providing a nontrivial consistency check: We calculated the scalar
product using $q$-analysis on a discrete $q$- Fock space representation.
However, the calculation \cite{ak-gm-rm-prd} of this scalar product
in the special case $\alpha =0$, which we here recover
in the limit, had been performed
with entirely different analytic methods in a continuous representation.
\section{Self-adjoint extensions of $\x$ and $\p$}
\label{inv}
In this section we continue formal considerations of
\cite{ak-jmp-ucr} where it was proved that
the operators $\x$ and $\p$ separately
do have one-parameter families of self-adjoint extensions in $H$.
To be precise, $\x$ on $D$ is symmetric, while its adjoint $\x^*$
is closed but nonsymmetric. $\x^{**}$ is closed and symmetric
and has deficiency indices (1,1).
There are families of diagonalisations of $\x$ in $H$, though
of course not in $D$. The same holds for $\p$.
The corresponding eigenvectors are unphysical states,
separated from the physical domain by the minimal uncertainty gap, see
Eq.\ref{gap}.
\sn
While in \cite{ak-jmp-ucr} the existence only of self-adjoint
extensions had been proven,
we can now explicitly solve the eigenvalue problem
\be
\x.\v{v_{x}} = x \v{v_{x}}
\ee
For the solution we make the ansatz
\be
\v{v_{x}} \;=\; N^{-1}(x) \sum_{n=0}^{\infty} q^{-n/2} d_n(x)\v{n}
\ee
yielding the recurrence relation
\be
\frac{x}{L} d_n(x) \;=\;
\sqrt{[n+1]q^{-1}}\, d_{n+1}(x) \,+\, \sqrt{q[n]} \, d_{n-1}(x)
\label{e80}
\ee
with $d_{-1} =0$ and $d_0 = 1$.
In fact, Eq.\ref{e80} can be transformed into the recurrence relation
Eq.\ref{reca}, i.e. the $d_n$ can be transformed into our previously
considered coefficients $c_n$:
\be
d_n(x) \;=\; c_n(2 x (q+q^{-1})^{-1})\;.
\ee
In the expansion of $\v{v_x}$, the
factor $q^{-n/2}$ is different from the corresponding factor
$q^{-3n/2}$ in the expansion of the $\v{\psi^{ml}_x}$, implying that
the scalar product and normalisation constant of the formal
eigenvectors are different from those of the
maximal localisation states which we had calculated earlier:
\be
N(x) \;=\;= \frac{q^{j^2 + 1/4}\, \theta_4(\pi j, \lambda^2)\,
 \theta_2(0,q^{-1})}{
(q^j+q^{-j})\, \theta_4(0,\lambda^2)}
\ee
\be
\sp{v_x}{v_{x^{\prime}}} = \frac{A^2 q^{1/2(j^2 + j^{\prime}{}^2+1)}
\theta_1(\frac{\pi}{2}(j-j^{\prime})\lambda^2)\,
\theta_4(\frac{\pi}{2}(j+j^{\prime}),\lambda^2)
}{(x-x^{\prime})\,\bar{\omega}\,\sqrt{N(x) N(x^{\prime})}}
\ee
where now
\be
x(j) := \frac{q^j -q^{-j}}{2 \bar{\omega}} \qquad \mbox{with} \qquad
\bar{\omega} = \frac{\sqrt{q^2-1}}{2L}
\ee
and where we abbreviated again $\index^\prime := \index(x^\prime)$.\newline
We draw the graph of the scalar product over $j$ for $j^{\prime}=0$:
\vskip5truecm
\epsfxsize=5in \centerline{\epsfbox{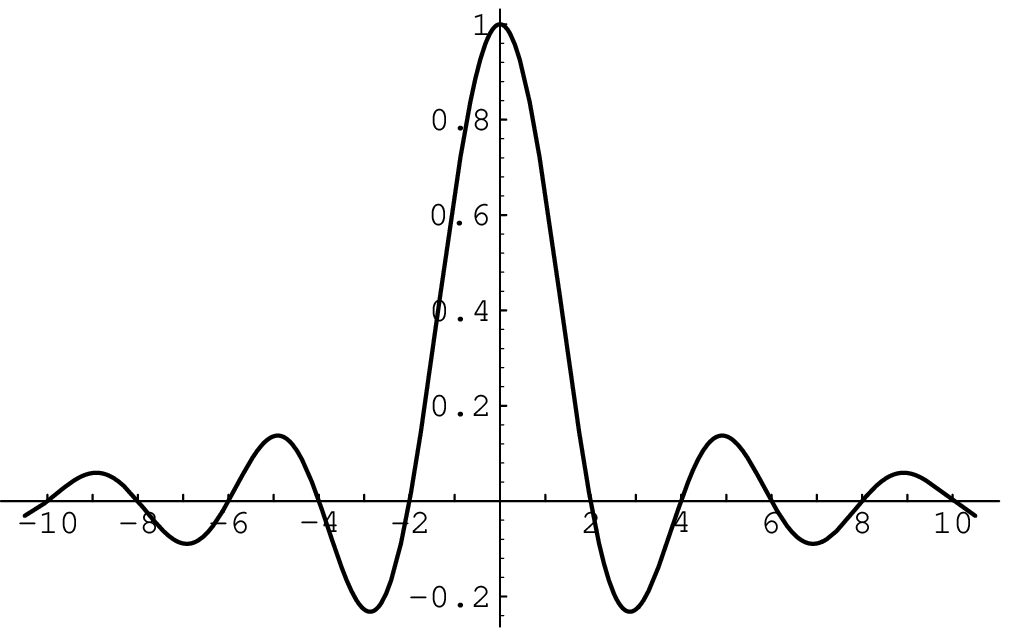}}
\centerline{{\small \it Fig. 2: Scalar product $\sp{v_x}{v_{0}}$
of formal eigenvectors drawn over $j(x)$}}
\vskip0.8cm
\noindent From the zero's of $\theta_1$
we read off that the $\v{v_x}$ are mutually othogonal for
$j-j^{\prime} \in 2 \N$. Using $j^{\prime}$ as a parameter in the
range $j^{\prime} \in [0,2[$ we identify for each value of $j^{\prime}$
a diagonalisation of $\x$.
Thus, $j^\prime$ labels the self-adjoint extensions with
the corresponding eigenvalues $(x_n)_{n\in \N}$ being
\be
x_n = \frac{q^{2n + j^{\prime}} - q^{-2n -j^{\prime}}}{2\sqrt{q^2-1}} L
   = \frac{\sinh((2n+j^{\prime})\ln q)}{\sqrt{q^2-1}} L \qquad (n\in \N)
\label{a61}
\ee
Compare also with the graph of the scalar product which had been calculated
only numerically in \cite{ak-jmp-ucr}. Having found the analytic
form of the scalar product in terms of $\theta$ functions we were able
to determine the one parameter familiy of diagonalisations of $\x$, of
which we had so far only known its existence.
As is not difficult to see we recover for vanishing minimal uncertainty in
momentum, i.e. for $\alpha \rightarrow 0$, the linear spectrum found in
\cite{ak-gm-rm-prd} for that special case.

Analogously to above we obtain the eigenvalues of $\p$ in its
self-adjoint extensions ($j^{\prime\prime} \in [0,2[$):
\be
p_n = \frac{q^{2n + j^{\prime\prime}} - q^{-2n
-j^{\prime\prime}}}{2\sqrt{q^2-1}} K
   = \frac{\sinh((2n+j^{\prime\prime})\ln q)}{\sqrt{q^2-1}}
   K \qquad (n\in \N)
\label{a62}
\ee
We stress that the parameters $j^\prime, j^{\prime\prime}$
of Eqs.\ref{a61},\ref{a62} label \it different \rm extensions of the
domain $D$ of $\x$ and $\p$. Recall that the uncertainty relation
implies that the formal $\x$- or $\p$- eigenvectors which we here calculated
do not lie in any \it common \rm extension of the
domain $D$ of $\x$ and $\p$. They are not physical
states and are separated from the physical domain by the uncertainty gap,
see Eq.\ref{gap}.
\sn
However, these families of diagonalisations of $\x$ or $\p$
in $H$- can still be of use, e.g. for the calculation of inverses
of $\x$ and $\p$, which would have been difficult to invert as nondiagonal
operators in the Fock basis $\x = L(a + a^{\dagger})$
and $\p = i K(a-a^\dagger)$.

\section{Outlook}
In quantum field theory, interaction terms which
on ordinary geometry would be ultraviolet regular but nonlocal,
can in fact be regular and strictly local on a geometry with a minimal
position uncertainty.
The reason is that an interaction is to be considered
strictly local if no nonlocality could be observed. Intuitivly
this is the case if a small apparent
nonlocality of the interaction term is unobservable due to a
comparatively larger minimal uncertainty in the underlying space.
We already mentioned that, as
has been shown in \cite{ak-np3}, quantum field theories can be
naturally regularised when working on a generalised
geometry with intrinsic minimal uncertainties.
Generally, in order to explicitly compare the size of nonlocality
of an arbitrary interaction term with the size of the intrinsic uncertainty
of the generalised geometry it is crucial to have available the states of
maximal localisation on this geometry. Similarly, maximal localisation
in a momentum space with minimal uncertainty $\Delta p_0$ is
of interest in the context of infrared regularisation.

So far, we have studied the properties of the
maximal localisation states in one dimension only.
The generalised Fourier transformations which map between
the (quasi-) position and the (quasi-) momentum representations have only
been studied in the special case $\alpha =0$.
For the general case techniques should be useful
which have been developed for the Fourier theory \cite{ak-sm-jmp}
on quantum planes \cite{FRT}.
Also, the unitary equivalence of \it all \rm Hilbert space
representations, in the sense in which it holds for the ordinary commutation
relations, has not yet been proven.
Most interesting further physical insight into the nature of these
generalised geometries can be expected from studies on maximal
localisation in $n$ dimensions where
$[\x_i,\x_j]\ne 0$ and $[\p_i,\p_j]\ne 0$
lead also to $\Delta x_i \Delta x_j \ne 0$ and $\Delta p_i \Delta p_j \ne 0$.
Work in this direction is in progress.
\bn
\bf Acknowledgement \rm

A.K. would like to thank Prof. T Koornwinder for the kind invitation to
give a seminar in Amsterdam and a very useful discussion on $q$- Hermite
polynomials.

\section{Appendix A}
%
%
In the following we outline a proof for the equivalence of the
formula of Rodriguez Eq.\ref{Rodriguez}
and the recurrence relation Eq.\ref{reca}.
We use the notation
$A f(x_\index) := \frac12\Bigl(f(x_{\index-1})+f(x_{\index+1})\Bigr)$
which allows the differentiation of products to be written
in the form:
\be
\label{ProductRule}
D(fg) \;=\; (Df)(Ag) + (Af)(Dg)\;.
\ee
We first prove the identity
\be
\label{AppEq1}
D^{n+1} q^{-j^2} \;=\; -\Bigl( \frac{q+q^{-1}}{2L} \Bigr)^2 \,
\Bigl( q^{-2n+1} [n]\,D^{n-1} + x\,q^{-n} D^n \Bigr) \, q^{-j^2}\;.
\ee
The case $n=0$ can be verified easily by hand.
Induction from $n$ to $n+1$ implies that
\be
\label{AppEq2}
D^{n+1} q^{-j^2} \;=\; D(D^n q^{-j^2}) \;=\;
 -D\,\Bigl( \frac{q+q^{-1}}{2L} \Bigr)^2 \,
\Bigl( q^{-2n+3} [n-1]\,D^{n-2} + x\,q^{-n+1} D^{n-1} \Bigr) \, q^{-j^2}
\ee
Comparing the r.h.s. of Eqs.\ref{AppEq1}
and \ref{AppEq2} and using Eq.\ref{ProductRule}, one is led to the condition
\be
\label{AppEq3}
\Bigl( A\,D^{n+1} + \frac12(q-q^{-1})\,x\,D^n - q^{-n}D^{n-1}
\Bigr) \, q^{-j^2} \;=\; 0
\ee
which can be proved by a further induction where the identities
$$
Dx=1\,,\hspace{8mm}
Ax=\frac12(q+q^{-1})x \,,\hspace{8mm}
DA-\frac12(q+q^{-1})AD=\frac14(q-q^{-1})^2x\,D^2
$$
turn out to be very useful. Once Eq.\ref{AppEq1}
is proved, one obtains the recursion relation
Eq.\ref{reca} by inserting the formula
of Rodriguez which completes the proof.
%
%
\section{Appendix B}
%
%
We prove the limits $c^\pm(x)$ i.e. we prove
Eqs.\ref{Cplus},\ref{ConstA}:
\be
\label{Lemma1}
c^+(x_\index) \;=\; \lim_{m \rightarrow \infty} \, (-)^m \, c_{2m}(x_\index)
\;=\;  \sqrt{\frac{\pi}{2 \,(q^{-2};q^{-2})_\infty^3\,\ln q }}
\,q^{\index^2/2}\,\theta_2 (\frac{\pi \index}{2},\lambda)
\ee
Let us first rewrite expression Eq.\ref{Cexplicit} by
\be
c_{2m} (x_\index) \;=\; \frac{1}{q^{2m^2} \,\sqrt{(q^{-2};q^{-2})_{2m}}} \,
\sum_{k=-m}^{m} \biggl( {2m \atop m+k} \biggr)_{q^2}
(-)^{m+k} \, q^{2k\index}\,.
\ee
Choosing an integer $0<r<m$ we split up this sum into two parts
\be
(-)^m \, c_{2m}(x_\index) = S_{m,r}^{(1)} + S_{m,r}^{(2)}
\ee
where
\ba
S_{m,r}^{(1)} &=& \frac{1}{q^{2m^2} \sqrt{(q^{-2};q^{-2})_{2m}}}
\sum_{k=-r}^{r}
\biggl( {2m \atop m+k} \biggr)_{q^2} (-)^k \, q^{2k\index} \\
S_{m,r}^{(2)} &=& \frac{1}{q^{2m^2} \sqrt{(q^{-2};q^{-2})_{2m}}}
\sum_{k=r+1}^{m}
\biggl( {2m \atop m+k} \biggr)_{q^2} (-)^k \,
(q^{2k\index}+q^{-2k\index})
\ea
Now let $m$ go to infinity and keep $r$ fixed. Since $q^2>1$ and
Eq.\ref{Qinverse} we have the identity
\be
\label{BinomIdentity}
\biggl( {2m \atop m+k} \biggr)_{q^2} \;=\;
q^{2 (m^2-k^2)} \biggl( {2m \atop m+k} \biggr)_{q^{-2}}\,.
\ee
Because of
\be
\lim_{m \rightarrow \infty} \,
\biggl( {2m \atop m+k} \biggr)_{q^{-2}}
\;=\; \frac{1}{(q^{-2};q^{-2})_\infty}
\ee
the first part converges to
\be
S_{r}^{(1)} \;=\;
\lim_{m \rightarrow \infty}S_{m,r}^{(1)} \;=\;
\frac{1}{(q^{-2};q^{-2})_\infty^{3/2}} \,
\sum_{k=-r}^r\,(-)^k\,q^{2k\index-2k^2}\,.
\ee
The second part $S_{m,r}^2$ can be estimated as follows. As can be
seen from Eq.\ref{BinomIdentity} the inequality
\be
\biggl( {2m \atop m+k} \biggr)_{q^{-2}} \;=\;
\frac{\prod_{i=m+k+1}^{2m} (1-q^{-2i})}{\prod_{i=1}^{m-k} (1-q^{-2i})}
\;\leq\; \frac{1}{\prod_{i=1}^{2m} (1-q^{-2i})} \;=\;
\frac{1}{(q^{-2}; q^{-2})_{2m}}
\ee
implies that
\be
\biggl( {2m \atop m+k} \biggr)_{q^2} \;\leq\;
\frac{q^{2(m^2-k^2)}}{(q^{-2};q^{-2})_\infty}\,.
\ee
Therefore
\be
|S_{m,r}^{(2)}| \leq
\frac{2}{q^{2m^2} \sqrt{(q^{-2};q^{-2})_{2m}}}
\sum_{k=r+1}^{m} \biggl( {2m \atop m+k} \biggr)_{q^2}
\, q^{2k\index}
\leq
\frac{2}{(q^{-2};q^{-2})_{2m}^{3/2}}
\sum_{k=r+1}^{m} \, q^{2(k\index-k^2)}
\ee
so that
\be
\label{Estimation}
|S_{r}^{(2)}| \;=\;
|\lim_{m \rightarrow \infty} \,S_{m,r}^{(2)}| \;\leq\;
\frac{2\,q^{2r\index-2r^2}}{(q^{-2};q^{-2})_{\infty}^{3/2}}
\sum_{k=1}^{\infty} \, q^{2(k\index-k^2)}\,.
\ee
Since the sum on the r.h.s. is finite this expression tends to
zero as $r$ goes to infinity. Thus we conclude that
\be
\label{SumExpression}
c^+(x_\index) \;=\; \lim_{r\rightarrow\infty} S_{r}^{(1)} \;=\;
\frac{1}{(q^{-2};q^{-2})_\infty^{3/2}} \,
\sum_{k=-\infty}^\infty\,(-)^k\,q^{2k\index-2k^2}\,.
\ee
The sum on the r.h.s. of this expression
is essentially a Jacobi $\theta_2$-function.
In order to see this notice that its definition Eq.\ref{Theta2}
can also be written as
\be
\theta_2(z,e^{- \tau}) \;=\; \sqrt{\frac{\pi}{\tau}}\,
\sum_{k=0}^{\infty} \, (-)^k \, e^{-\frac{1}{\tau}(z-\pi k)^2}\,.
\ee
Inserting $\tau=- \ln \lambda = \frac{\pi^2}{2 \ln q}$ and
$z=\frac{\pi \index}{2}$ we can express the sum in
Eq.\ref{SumExpression} by
\be
\sum_{k=-\infty}^{\infty} \, (-)^k \, q^{-2k^2+2k\index} \;=\;
\sqrt{\frac{\pi}{2 \ln q}} \, q^{\index^2/2} \, \theta_2(z,\lambda)
\ee
which completes the proof for $c^+(x_\index)$. The proof for
$c^-(x_\index)$ follows the same lines.

\end{document}